\documentclass[aps,prl,twocolumn,showpacs,superscriptaddress,longbibliography]{revtex4-1}
\usepackage{amsbsy,amssymb,amsmath,bm}
\usepackage{graphicx}
\usepackage{braket}
\usepackage{float}
\usepackage{textcomp}
\usepackage{color}
\usepackage[colorlinks=true,linkcolor=red]{hyperref}
\usepackage[sort&compress]{natbib}

\usepackage{soul}

\input{epsf}

\begin{document}

\title {
Universal Nonlinear Disordered Wave Packet Subdiffusion: 12 Decades
}
\author{Ihor Vakulchyk}
\affiliation {Center for Theoretical Physics of Complex Systems, Institute for Basic Science(IBS), Daejeon, Korea, 34126}
\affiliation{Basic Science Program, Korea University of Science and Technology(UST),  Daejeon, Korea, 34113}
\author{Mikhail V. Fistul}
\affiliation {Center for Theoretical Physics of Complex Systems, Institute for Basic Science(IBS), Daejeon, Korea, 34126}
\affiliation{Russian Quantum Center, National University of Science and Technology "MISIS", 119049 Moscow, Russia}
\author{Sergej Flach}
\affiliation {Center for Theoretical Physics of Complex Systems, Institute for Basic Science(IBS), Daejeon, Korea, 34126}

\date{\today}

\begin{abstract}
We use a novel unitary map toolbox -- discrete time quantum walks originally designed for quantum computing -- to implement ultrafast
computer simulations of extremely slow dynamics in a nonlinear and disordered medium. Previous reports on wave packet spreading in Gross-Pitaevskii lattices observed subdiffusion with the second moment $m_2 \sim t^{1/3}$ (with time in units of a  characteristic scale $t_0$) 
up to the largest computed times of the order of $10^8$. A fundamental question remained as to whether this process can continue ad infinitum, or has to slow down. Current experimental devices are not capable to even reach 1\textpertenthousand ~of the reported computational horizons. With our toolbox, we outperform previous computational results and observe that the universal subdiffusion persists over additional four decades reaching 'astronomic' times $2\cdot 10^{12}$.
Such a dramatic extension of previous computational horizons suggests that subdiffusion is universal, and
that the toolbox can be efficiently used to assess other hard computational many-body problems.

\end{abstract}

\maketitle

Eigenstates of linear excitations in a one-dimensional medium exposed to an uncorrelated random external field are exponentially localized in space, due to the celebrated Anderson localization (AL) \cite{anderson1958absence}. 
Thus any evolving compact wave packet in such a system will first spread, but then halt and not escape from its localization volume.  The width of the wave packet will be of the order of the localization length $\xi$ \cite{lifshits1988introduction}.
Experimental verifications of AL with Bose-Einstein condensates of ultracold atomic gases loaded onto optical potentials were using precisely the above technique, i.e. the time evolution of a wave packet, to prove and quantitatively characterize the degree of AL \cite{Billy:2008aa,Roati:2008aa}.  Numerous further experimental studies of AL employ
light \cite{Schwartz:2007aa,lahini2008anderson}, microwaves \cite{Dalichaouch:1991aa}, and ultrasound \cite{WEAVER1990129,Hu:2008aa}, among
others
(see also Ref.\cite{1751-8121-47-49-493001} for a recent review).

The interplay of disorder with many body interactions intrigued the minds of researchers ever since AL was established.
Recent experimental attempts include granular chains \cite{kim2018direct}, photonic waveguide lattices \cite{lahini2008anderson}, light propagation in fiber arrays \cite{pertsch2004nonlinearity},
and atomic Bose-Einstein condensates \cite{lucioni2011observation}. 
In particular, the latter case studying the spatial extension of clouds of interacting $^{39}{\rm K}$ atoms revealed the destruction of AL through the onset of subdiffusion -- an extremely slow process of wave packet spreading with its second moment $m_2 \sim t^{\alpha}$ with $\alpha < 1$. Here time $t$ is measured in units of a characteristic microscopic time scale $t_0$. 
E.g. for ultracold atomic gases this time scale $t_0 \simeq 1ms$ \cite{lucioni2011observation}. Different values of the exponent $\alpha$ were measured, which ranged between $0.1$ and $0.5$. That imprecision is due to the slow dynamics of subdiffusion that did not allow to quantitatively assess the subdiffusion exponents.  
E.g. in the atomic gas case, the need
to keep the condensate coherent, results in a  time limitation of about $10^4$, i.e. about 10s. 
Thus there is clear need of alternative computational studies which may shed light on the fate of expanding wave packets. With the large number ($\sim 10^6$) of interacting cold atoms, semiclassical approximations lead to effective nonlinear wave equations similar to the celebrated Gross-Pitaevskii one, with two-body interactions turning into quartic anharmonicity.

Computational studies of spreading wave packets in various nonlinear and disordered systems revealed interesting features. 
On times up to $10^2$, an initially compact wave packet expands up to the size of localization length $\xi$. After that  a subdiffusive spreading of the wave packet \cite{PhysRevLett.100.094101,flach2009universal} with $\alpha \approx 1/3$ occurs. 
At large times the wave packet is composed of a still growing central flat region of size  $m_2^{1/2} \gg \xi$ with sharp boundaries of size $\xi$. 

Qualitatively the subdiffusion can be explained as follows. The chaotic dynamics inside the wave packet leads to dephasing of the participating localized Anderson normal modes. With coherence lost, wave localization cannot be anymore sustained. The assumption of strong chaos (quick and complete dephasing of \textit{all} modes) results in the stochastic interaction between the localized normal modes and the prediction $\alpha=1/2$ \cite{flach2009universal}. An additional phenomenological estimate of the impact of \textit{finite but small probabilities of resonances between interacting normal modes }
finally leads to a substantial suppression of dephasing and the correct value  $\alpha=1/3$ \cite{flach2009universal}.
Interestingly the validity of this estimate was confirmed with tests of its predictions for larger system dimensions \cite{0295-5075-98-6-60002}, and different exponents of nonlinear terms which correspond to various $N$-body interactions
\cite{PhysRevE.82.016208}. Successful tests of systems with quasiperiodic (instead of random) potentials \cite{1367-2630-14-10-103036}, and nonlinear versions of quantum kicked rotors \cite{PhysRevLett.70.1787,gligoric2011interactions} yielded subdiffusion with $\alpha=1/3$ as well, and revealed additional universality aspects of the
observed process \cite{1751-8121-47-49-493001}. 
The largest times reached by these computations were $10^8-10^9$.  In the case of one single disorder realization a reported
evolution for a Klein-Gordon chain reached time $10^{10}$ \cite{PhysRevE.79.056211}. As a side note, 
in weakly nonlinear systems
Anderson
localization in the evolution of finite size wave packets is restored in a
probabilistic manner
\cite{johansson2010kam,PhysRevLett.107.240602,basko2011weak}.

To address the fundamental question whether wave packet spreading slows down or continues, we use a novel unitary map toolbox -- Discrete Time Quantum Walks (DTQW). We peek beyond
previous horizons set by the CPU time limits for systems of coupled ordinary differential equations. We obtain results for unprecedented times up to $2 \cdot 10^{12}$ and thereby shift the old Gross-Pitaevskii horizons by four decades. 

DTQW were introduced as quantum generalizations of classical random walks by Aharonov et al. \cite{aharonov1993quantum}. 
The DTQW evolution is a (discrete) sequence of unitary operators acting on a quantum state in a high dimensional Hilbert space. DTQW shows quantum interference/superposition \cite{aharonov1993quantum}, entanglement \cite{abal2006quantum}, two-body coupling of wave functions \cite{omar2006quantum}, Anderson localization \cite{crespi2013anderson}, etc. 
DTQW experimental realizations were reported with ion trap systems \cite{schmitz2009quantum}, quantum optical waveguides \cite{peruzzo2010quantum}, and nuclear magnetic resonance quantum computer \cite{du2003experimental}. Quantum walks are studied by the quantum computing community since they allow for implementing algorithms which exponentially surpass known classical ones \cite{childs2003exponential, farhi2008quantum}.

\begin{figure}[tbp]
\includegraphics[width=0.95 \columnwidth]{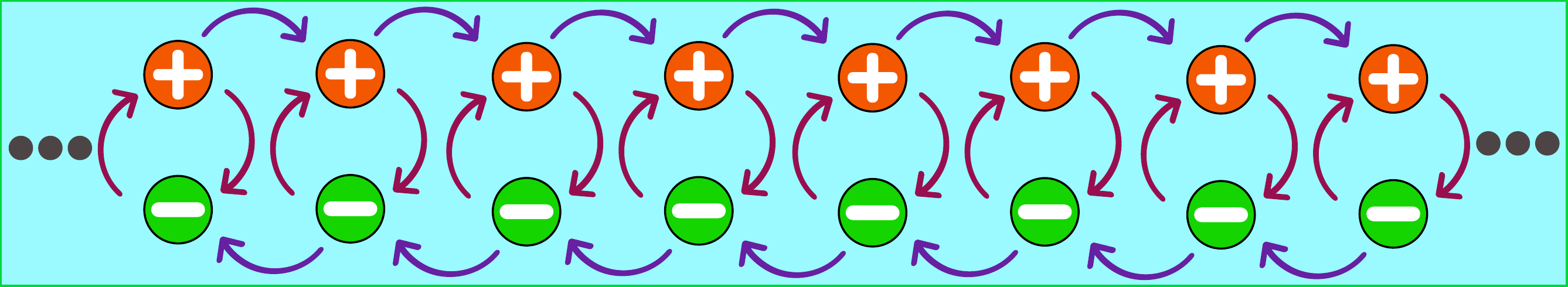}
\caption{ A schematic representation of a general discrete-time quantum walk. The vertical arrows indicate the quantum coin 
action within each two level system, while the horizontal ones
show the action of the transfer operator.
}
\label{fig1}
\end{figure}

Consider a single quantum particle with an internal spin degree of freedom, moving on a one-dimensional lattice. Its dynamics is determined by a time- and lattice-site-dependent two-component wave function $\Psi_n (t)=\{\Psi_n^+,\Psi_n^- \}$, which evolves under the influence of some periodic Floquet drive. Then, its evolution can be mapped onto a sequence of unitary maps. 
The unitary map evolves the wave function over one period of the unspecified Floquet drive. For a DTQW, this map is the product of a {\sl coin} operator $\hat S$ and a {\sl transfer} operator $\hat{T}_{\pm}$.
The coin operator $ \hat S = \sum_n \hat{S}_n= \sum_n \hat{V}_n \ket{n}\bra{n}$, with the single site quantum coin 
\begin{equation}\label{coin_operator}
    \hat V_n =
        \begin{pmatrix}
        \cos{\theta} & e^{i \varphi_n} \sin{\theta} \\
        -e^{-i \varphi_n} \sin{\theta} &  \cos{\theta}
        \end{pmatrix}. 
\end{equation}
A schematic map flow is shown in Fig.\ref{fig1}.
This unitary matrix parametrization is a particular realization of the general case discussed in Ref. \cite{vakulchyk2017anderson}, with two angles $\theta$ (kinetic energy) and $\varphi_n$ (site dependent
internal synthetic flux).
Such coin operators can be implemented by time--dependent perturbations \cite{CoinOperImpl,Impl1,di2004cavity,Impl3}. 

The transfer operator
\begin{equation}\label{shift_operator}
  \hat T_\pm = \sum_n \ket{n}\bra{n+1} \otimes \ket{\mp}\bra{\mp} +
  \ket{n}\bra{n-1} \otimes \ket{\pm}\bra{\pm},
\end{equation}
with $\pm$ corresponding to the $+$ components shifting either to the right  or to the left. We will use $T_+$ across the paper. The DTQW evolution follows as a sequence of successive $\hat S$ and $\hat T_+$ operators acting on the state:
\begin{equation} \label{HoppingEquation}
\Psi_n(t+1)= \hat{M}_{n-1,+} \Psi_{n-1}(t)+ \hat{M}_{n+1,-} \Psi_{n+1}(t),
\end{equation}
where the matrices $\hat{M}_{n,\pm}$ are defined by the elements of $\hat{V}_n$:
\begin{equation} \label{Mmatrix+}
\hat{M}_{+}= \left ( \begin{array}{cc}
V_{11}&   V_{12} \\
0& 0\\
\end{array} \right )
\;,\;
\hat{M}_{-}= \left ( \begin{array}{cc}
0&  0\\
V_{21} & V_{22} 
\end{array} \right )\;.
\end{equation}

\begin{figure}[tbp]
\includegraphics[width=0.95 \columnwidth]{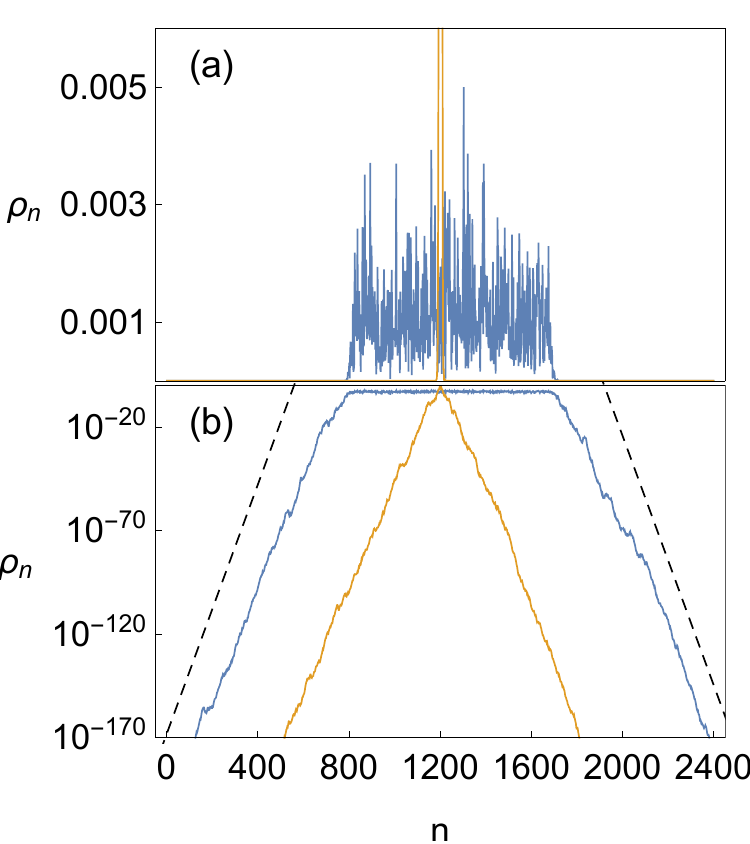}
\caption{
Wave packet density profiles $\rho_n$ for linear $g=0$ (orange solid lines) and nonlinear $g=3$ (blue solid lines) DTQWs at the time $t_f=10^8$ (linear case) and $t_f=2 \times 10^{12}$ (nonlinear case). 
(a) lin-lin plot.
(b) log-normal plot. 
The black dashed lines indicate exponential decay with the corresponding localization length ${\rm e}^{-2|n-n_0|/\xi}$.  
}
\label{fig2}
\end{figure}
Next, we consider a strongly disordered
DTQW with random uncorrelated angles $\varphi_n$ being
uniformly distributed over the entire existence domain $[-\pi, \pi)$. The resulting unitary
eigenvalue problem is solved by finding the orthonormal set of eigenvectors
$\{  \tilde{\Psi}_{\nu,n} \}$ with
$\tilde{\Psi}_{\nu,n}(t+1)=e^{-i\omega_{\nu}  } \tilde{\Psi}_{\nu,n}(t)$ and the eigenvalues $e^{-i\omega_{\nu}}$, where $\omega_{\nu}$ is the quasienergy.
All eigenvectors are exponentially localized on the chain \cite{vakulchyk2017anderson}. This is a manifestation of Anderson localization.
Remarkably, for such strong disorder all eigenvectors $\tilde{\Psi}_{\nu,|n| \rightarrow \infty} \sim e^{-|n|/\xi}$ 
are characterized by
{\sl one single localization length} $\xi(\theta)$ which {\sl does not} depend on the quasienergy $\omega_{\nu}$ of a given state 
\cite{vakulchyk2017anderson}:
\begin{equation}\label{loc_length}
 \xi = -\frac{1}{\ln(\left| \cos(\theta) \right|)}.
\end{equation}
Another remarkable feature is that for any value of the localization length -- either small or large compared to the
lattice spacing $\Delta n \equiv 1$ -- the spectrum of the quasienergies of an infinite chain is densely filling
the compact space of angles of complex numbers on a unit circle \cite{vakulchyk2017anderson}. 
Therefore, the density of states is constant, and gaps in the spectrum are absent.

\begin{figure}[tbp]
\includegraphics[width=0.95 \columnwidth]{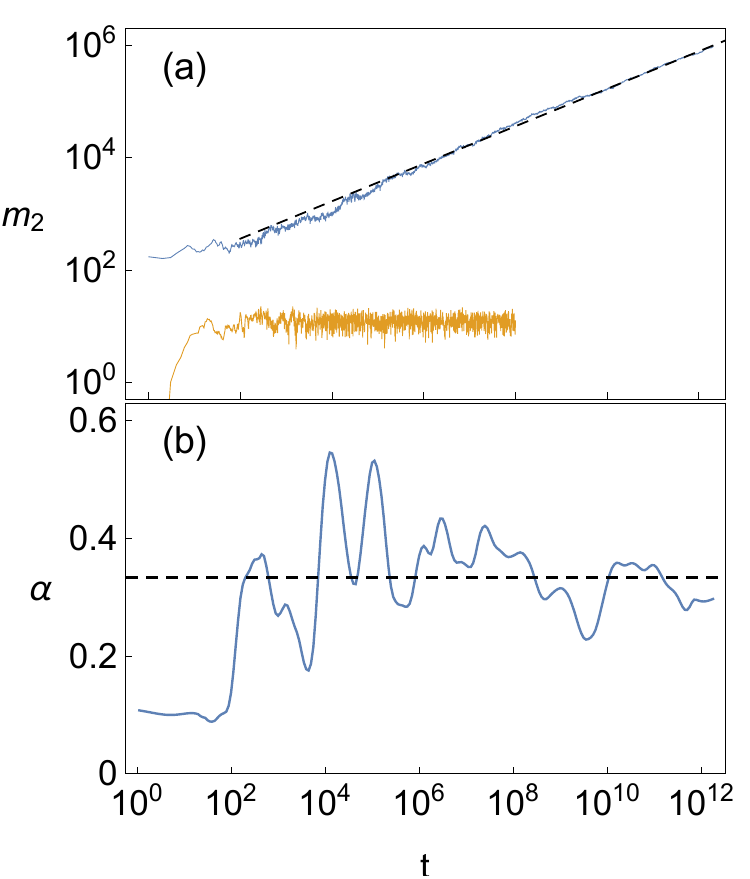}
\caption{(a) $m_2(t)$ versus time in log-log scale for the linear
$g=0$ (orange solid lines) and nonlinear $g=3$ (blue sold lines) case from Fig.\ref{fig2}.
(b) The derivative $\alpha(t)$ versus time for the data of the nonlienar run from (a) (blue line). Black dashed line corresponds to $\alpha=1/3$.
\label{fig3}}
\end{figure}

Anderson localization is manifested through the halt of spreading of an evolving wave packet. In our direct numerical simulations, we choose $\theta=\pi/4$, which results in a localization length $\xi \approx 2.9 $ and a typical localized Anderson 
eigenstate occupying about 10 lattice sites.
We choose the initial state to be localized on $M$ sites:
\begin{equation}
\label{IC}
\Psi_n(t=0)=\frac{1}{\sqrt{2M}}(1,{\rm i})\;,\;n=n_0, ..., (n_0+M-1)   \;.
\end{equation}
We evolve this state using Eq.(\ref{HoppingEquation}) until $t=10^8$ for a system of size $N=2400$ and $M=1$. The density distribution $\rho_n(t) = |\Psi_n^-|^2 + |\Psi_n^+|^2$ observed for such time is presented in  Fig. \ref{fig2}(a) (orange solid lines).
The distribution is clearly localized, with the width of a few localization lengths.
The tails are exponentially decaying, with a slope which is well fitted using the localization length $\xi$ in Fig. \ref{fig2}(b).  
To further quantify the halt  of spreading, we compute the first moment $\bar{n}(t) = \sum_{n=1}^N n \rho_n(t)$, and
then the central object of our studies -- the second moment
\begin{equation}
    m_2(t) = \sum_{n=1}^N (n-\bar{n}(t))^2 \rho_n(t) \;,
\label{m2}
\end{equation}
The time dependence of $m_2$  is plotted in Fig.\ref{fig3}(a). We observe a halt of the growth of $m_2(t)$ at $t\approx 10^2$, which 
together with the profile of the halted wave packet (see Fig. \ref{fig2}) is a clear demonstration of Anderson localization.

We now leave the grounds of linear DTQW and generalize the DTQW to a {\sl nonlinear} unitary map by adding a density-dependent renormalization to the angle 
\begin{equation}
\label{nlangle}
    \varphi_{n} = \xi_n + g \rho_n\; ,
\end{equation}
where $g$ is the nonlinearity strength. 
We note that (\ref{nlangle})  keeps the conservation of the total norm: $\sum_n \dot{\rho}_n=0$.

Our main goal is to measure the details of subdiffusive wave packet spreading on large time scales with $\rho_n \ll 1$.
Therefore we  use a low-density approximation for the  quantum coin (\ref{coin_operator}) which approximates the exponential factors of the coin without violating evolution unitarity:
\begin{equation}\label{sqrt_coin_model}
  e^{i \phi_n} = e^{i \xi_n} \left( \sqrt{1-g^2 \rho_n ^2} + i g \rho_n \right).
\end{equation}
The computational advantage of fast 
calculations of square roots as opposed to slow ones of trigonometric functions serves the purpose to further extend the 
simulation times. To guarantee unitarity of the evolution, we choose $M > 1$ such that $g^2 \rho_n^2 \ll 1$.  

We evolve a wave packet with $g=3$ and $M=13$ and plot the density distribution at the final time $t_f=2\cdot 10^{12}$  in Fig.\ref{fig2} (blue solid line). This is a new record evolution time, beating old horizons by a factor of $10^4$. We observe a familiar structure of the wave packet: a homogeneous wide central part with clean remnants of Anderson localization in the tails (Fig. \ref{fig2}(b)). The width of the wave packet reaches about 900 sites and exceeds the localization length $\xi$   by a stunning factor of about $300$. The time dependence of the second moment $m_2(t)$ is plotted in Fig.\ref{fig3}(a). A clean and steady growth of the wave packet width is evident, and the linear fitting on the log-log scale  (black line in Fig.  \ref{fig3}(a)) indicates the universal $\alpha=1/3$ value.  

However, we note that a straightforward fitting with a single power law can yield misleading results, since it is not evident where the asymptotic regime (if any) will start. To study the asymptotic regime in detail we quantitatively assess it by applying standard methods of simulations and data analysis \cite{supp}. We calculate local derivatives on log-log scales to obtain  a time-dependent exponent $\alpha(t) = {\rm d}\left[ \ln(m_2)\right] / {\rm d}\left[ \ln t \right]$. 
The resulting curve is plotted in Fig.\ref{fig3}(b) and strongly fluctuates around the value 1/3. 

\begin{figure}[tbp]
\includegraphics[width=0.95 \columnwidth]{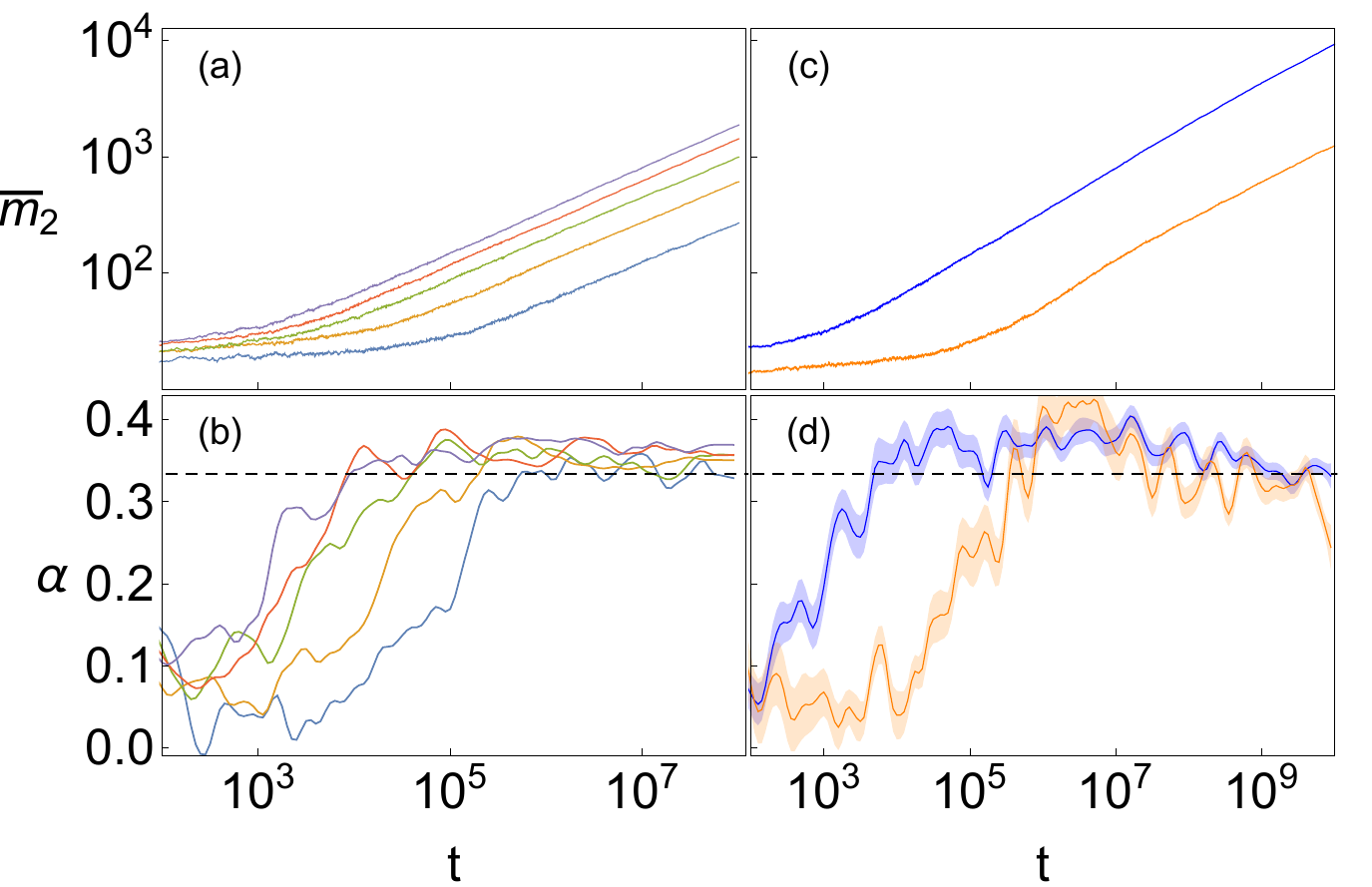}
\caption{ (a) The geometric average $\overline{m}_2$ versus time for 108 disorder realizations up to time $t=10^8$. 
The nonlinear parameter $g$ varies from bottom to top as $0.5$ (blue), $1$ (orange), $1.5$ (green), $2$ (red), $2.5$ (purple).
(b) The derivative $\alpha(t)$ versus time for the data from (a) and same color codes.
The horizontal dashed line corresponds to $\alpha=1/3$.
(c),(d) Same as in (a),(b) but for $g=2.5$ (blue) and $g=0.5$ (orange) and up to time $t=10^{10}$. Shaded areas indicate the statistical error.
}
\label{fig4}
\end{figure}

In order to reduce the fluctuation amplitudes, we obtain $m_{2,n}(t)$ for $R=108$ disorder realizations and compute
the geometric average $\ln \overline{m}_2(t) = \sum_n \ln m_{2,n}(t)/R$.
In Fig.\ref{fig4}(a) the results are shown for various values of $g=0.5,1,1.5,2,2.5$ up to times $10^8$ (the corresponding 
values of $M$ are $5, 8, 8, 10, 10$).  All curves 
approach the vicinity of $\alpha=1/3$ with fluctuation amplitudes substantially reduced as compared to the single run in 
Fig.\ref{fig3}(b). For $g=0.5$ and $g=2.5$ we extend the simulations up to time $10^{10}$ in Fig.\ref{fig4}(c) and observe a clear saturation 
of $\alpha(t)$ around 1/3 in Fig.\ref{fig4}(d). In particular, the weakest nonlinearity value $g=0.5$ is expected to 
show the earliest onset of asymptotic subdiffusion. Indeed we find in this case $\alpha =1/3 \pm 0.04$ starting with times
$t \geq 10^7$.

It is instructive to rewrite the evolution equations  in the basis of linear eigenmodes of the $g=0$ case. Using the Taylor expansion of nonlinear terms valid at large times when  $\rho_n \ll 1$ \cite{supp} 
we rewrite the unitary evolution of the wave function in the $g=0$ eigenmode basis
$\Psi_n(t) = \sum_{\alpha}  a_\alpha(t) \tilde{\Psi}_{\alpha,n}$:
\begin{eqnarray}
a_\alpha(t+1) = a_\alpha(t) e^{i \omega_\alpha} 
\nonumber
\\
+ i g \sin \theta \sum_{\alpha_1,\alpha_2,\alpha_3}^{2N}I_{\alpha,\alpha_1,\alpha_2,\alpha_3} a_{\alpha_1}(t)^*a_{\alpha_2}(t)a_{\alpha_3}(t),
\label{normal_modes_evolution}
\end{eqnarray}

with the overlap integral
\begin{equation}\label{interaction_integral}
    I_{\alpha,\alpha_1,\alpha_2,\alpha_3} = \sum_{k=1}^N \braket{\tilde{\Psi}_{\alpha,k}|\hat{u}^{(k)} \tilde{\Psi}_{\alpha_3,k}}\braket{\tilde{\Psi}_{\alpha_1,k} | \tilde{\Psi}_{\alpha_2,k}} .
\end{equation}
The structure of these resulting equations is strikingly similar to the ones obtained from Hamiltonian dynamics
\cite{PhysRevLett.100.094101,flach2009universal}. In particular, we obtain cubic nonlinear terms on teh rhs of the asymptotic
expansion (\ref{normal_modes_evolution}). Together with the one-dimensionality of the system the prediction of
a subdiffusive exponent $\alpha=1/3$ follows from Ref.\cite{PhysRevLett.100.094101,flach2009universal}.

To conclude, DTQW are very useful unitary map toolboxes which allow for extremely fast quantum evolution,
in particular due to covering finite times with one step (jump), and due to the fast(est) realization of a transfer/hooping/interaction
on a lattice.
We used a disordered version to obtain Anderson localization with the spectrum being dense and gapless and of compact support. The resulting localization length $\xi$ is not depending on the eigenvalue of an eigenstate and can be smoothly
changed in its whole range of existence using one of the control parameters of the DTQW. All these features induce
highest aesthetical satisfaction for what is to come. We then generalize the map to a nonlinear disordered DTQW and
study destruction of Anderson localization. Wave packets spread subdiffusively with their second moment $m_2 \sim t^{\alpha}$
and the universal exponent $\alpha=1/3$. The record time $t_f=2 \cdot 10^{12}$ is reached, which exceeds old horizons by
3-4 orders of magnitude. The size of the wave packet reaches $\approx 300 \xi$. The relative strength (or better weakness)
of the nonlinear terms in the DTQW reaches $0.01/\pi \approx 0.003$. No slowing down of the subdiffusive process was observed. Therefore chaotic dynamics appears to survive in the asymptotic limit of decreasing wave packet densities.

We expect DTQW to be useful in the future also for exploring other hard computational tasks, e.g. subdiffusion in two-dimensional and even three-dimensional nonlinear disordered lattices, and many body localization in interacting quantum
settings. 

\begin{acknowledgments}

This work was supported by the Institute for Basic Science, Project Code (IBS-R024-D1).
\end{acknowledgments}

\section{Supplementary Material}

\subsection{Asymptotic evolution equation}\label{sec3}

The discrete-time evolution is defined by a nonlinear map operator,
\begin{equation}\label{basic_evolution_map}
\begin{split}
      \ket{\Psi(t)} &= \hat{U}^t \ket{\Psi(0)}, \; t \in \mathbb{N} \\
    \hat{U} &= \hat{T} \otimes \sum_{n=1}^N \hat{S}_n.
\end{split}
\end{equation}
In order to separate the nonlinear components of $\hat{U}$ in $|\Psi_n| \rightarrow 0$, we consider a single coin operator on the site $n$. It can be factorized,
\begin{equation}
\begin{split}
    \hat{V}_n &= \hat{Z}_n \hat{V}^{(0)}_n \hat{Z}_n^{-1}, \\
    \hat{Z}_n =& \begin{pmatrix}
         \sqrt{1-g^2 \rho_n^2} + i g \rho_n  & 0 \\
        0 & 1
    \end{pmatrix},
\end{split}
\end{equation}
where $\hat{V}^{(0)}$ is the local coin operator under zero nonlinearity. Evaluating this to separate the linear part and consecutive nonlinear terms with different nonlinear exponent yields,
\begin{equation}\label{coin_matrix}
\begin{split}
    &\hat{V}_n = \hat{V}_n^{(0)} + i g \rho_n \sin{\theta} \begin{pmatrix}
        0 & e^{i\xi_n} \\
        e^{-i \xi_n} & 0 
    \end{pmatrix} \\
    + \sin{\theta} & \left( \sqrt{1 - g^2 \rho_n^2} -1 \right) \begin{pmatrix}
        0 & e^{i\xi_n} \\
        -e^{-i \xi_n} & 0 
    \end{pmatrix}.
\end{split}
\end{equation}
Using a Taylor expansion in the small nonlinearity limit, the evolution operator contains a sum of a linear term, a second-order (in $\left| \Psi_n \right|$) term and a sequence of $4k$-order terms $k \in \mathbb{N}$.  

The full non-linear map reads
\begin{equation}
\begin{split}
     \hat{U} &= \hat{U}^{(0)} + i g \sin{\theta} \sum_{k=1}^{N} \rho_k  \hat{u}^{(k)}  + \mathcal{O}\left(g^2 \rho_k^2\right),\\
      \hat{u}^{(k)} &= \hat{T} \otimes \left(  \begin{pmatrix}
        0 & e^{i \xi_n} \\
        e^{-i \xi_n} & 0
    \end{pmatrix}
     \ket{n} \bra{n} \right).
\end{split}
\end{equation}
We will explicitly evaluate the first-order non-negligible term in nonlinearity only. 

Let us consider the evolution in the linear limit eigenmode basis, 
\begin{equation}
    \hat{U}^{(0)} \ket{\tilde{\Psi}_{\alpha,k}} =  e^{i \omega_\alpha} \ket{\tilde{\Psi}_{\alpha,k}}, \quad \alpha = 1, 2, \ldots 2N. 
\end{equation}
With the wave function expanded in this basis $\Psi_{k}(t) = \sum_{\alpha} a_\alpha (t) \tilde{\Psi}_{\alpha,k}$, the evolution equations read
\begin{eqnarray}
    a_\alpha(t+1) = a_\alpha(t) e^{i \omega_\alpha}
\nonumber
\\
\label{normal_modes_evolution}
 + i g \sin \theta \sum_{\alpha_1,\alpha_2,\alpha_3}^{2N}I_{\alpha,\alpha_1,\alpha_2,\alpha_3} a_{\alpha_1}(t)^*a_{\alpha_2}(t)a_{\alpha_3}(t),
\end{eqnarray}
with the overlap integral (or matrix element, or overalp integral)
\begin{equation}\label{interaction_integral}
    I_{\alpha,\alpha_1,\alpha_2,\alpha_3} = \sum_{k=1}^N \bra{\tilde{\Psi}_{\alpha,k} }\hat{u}^{(k)}\ket{\tilde{\Psi}_{\alpha_3,k}}\braket{\tilde{\Psi}_{\alpha_1,k} | \tilde{\Psi}_{\alpha_2,k}}.
\end{equation}

\subsection{Numerical Approach}\label{sec4}

Let us discuss the details of simulations and data analysis. We directly propagate evolution equation (\ref{basic_evolution_map}). The initial conditions are uniformly spread over several neighboring sites to guarantee positivity and unitarity of the coins. 

The only source of the numerical error is the round-off errors of the finite dimensional computer algebra. We estimate the error by means of the total packet norm. It is equal to $1$ for initial conditions and supposed to be constant due to unitary evolution. The relative value of the error never exceeded $10^{-4}$.

In all the calculations we use $\theta=\pi/4$. The system size $N$ is between $2000$ and $2500$. The results which include ensemble averaging employ around $10^2$ realizations of the random field $\{\xi_n\}$. The total evolution times reach up to $2 \cdot 10^{12}$ time steps, which exceeds the maximum previously gained limits for such analysis to the best of our knowledge.    

The averaged curve of the second moment $m_2(t)$ is smoothed with the locally weighted regression smoothing (LOESS) \cite{cleavland1981lowess,william1988Locally} algorithm. The power-law exponent is then calculated as the two-point derivative of the smoothed data. To verify the smoothing procedure and exclude overfitting we also performed fitting with Hodrick-Prescott filter \cite{hodrick1997postwar}, Gaussian convolution smoothing and local 4th order polynomial fitting with an analytical derivative. We found all methods generating results within the statistical margin of error. The LOESS approach turned out to be more robust than other options. When averaging we estimate the error as the standard error of the mean.

To additionally speed up the simulations and reach further time limits we used GPU computing with CUDA language.

\bibliography{QW}

\end{document}